\documentclass[%
 aps,
 ph,%
 amsmath,amssymb,
 reprint,%
]{revtex4-1}
\usepackage[utf8]{inputenc}
\usepackage[T1]{fontenc}
\usepackage[colorlinks,citecolor=blue,urlcolor=blue,linkcolor=blue]{hyperref}
\usepackage{graphicx}
\usepackage{dcolumn}
\usepackage{bm}

\usepackage{multirow}

\usepackage{dirtytalk}

\usepackage{stmaryrd}

\begin{document}


\title[]{Diagnosis of two evaluation paths to density-based descriptors of \\ molecular electronic transitions}%

\author{Gabriel Breuil}
\email{gabriel.breuil@umontpellier.fr
}%
\affiliation{%
ICGM, University of Montpellier, CNRS, ENSCM, Montpellier, France%
}%

\author{Kaltrina Shehu}
\affiliation{%
ICGM, University of Montpellier, CNRS, ENSCM, Montpellier, France%
}%

\author{Elise Lognon}
\affiliation{%
ICGM, University of Montpellier, CNRS, ENSCM, Montpellier, France%
}%

\author{Sylvain Pitié}
\affiliation{%
ICGM, University of Montpellier, CNRS, ENSCM, Montpellier, France%
}%

\author{Benjamin Lasorne}
\affiliation{%
 ICGM, University of Montpellier, CNRS, ENSCM, Montpellier, France%
}%

\author{Thibaud Etienne}
\email{thibaud.etienne@umontpellier.fr
}%
\affiliation{%
ICGM, University of Montpellier, CNRS, ENSCM, Montpellier, France%
}%


\begin{abstract}
\noindent In this paper we discuss the reliability of two computational methods (numerical integration on Cartesian grids, and population analysis) used for evaluating scalar quantities related to the nature of electronic transitions. These descriptors are integrals of charge density functions built from the detachment and attachment density matrices projected in the Euclidean space using a finite basis of orbitals. While the numerical integration on Cartesian grids is easily considered to be converged for medium-sized density grids, the population analysis approximation to the numerical integration values is diagnosed using eight diagnostic tests performed on fifty-nine molecules with a combination of fifteen Gaussian basis sets and six exchange-correlation functionals. \\ $\;$ \\  \textit{Keywords: Excited states ; electronic-structure reorganization ; one-body reduced density matrices ; orbital population analysis.}

\end{abstract}

\maketitle

\section{Introduction}

\noindent The possibility of analyzing the charge transfer character of electronic transitions of complex molecular systems has been of interest for decades, and still attracts considerable attention in the excited-state electronic structure community \cite{luzanov_interpretation_1980,
head-gordon_analysis_1995,
furche_density_2001,
tretiak_density_2002,
martin_natural_2003,
batista_natural_2004,
tretiak_exciton_2005,
dreuw_single-reference_2005,
mayer_using_2007,
surjan_natural_2007,
wu_exciton_2008,
luzanov_electron_2009,
li_time-dependent_2011,
plasser_analysis_2012,
plasser_new_2014,
plasser_new_2014-1,
bappler_exciton_2014,
ronca_charge-displacement_2014,
ronca_density_2014,
etienne_toward_2014,
etienne_new_2014,
etienne_probing_2015,
pastore_unveiling_2017,
pluhar_visualizing_2018,
mewes_communication_2015,
li_particlehole_2015,
etienne_transition_2015,
plasser_statistical_2015,
poidevin_truncated_2016,
wenzel_physical_2016,
plasser_entanglement_2016,
savarese_metrics_2017,
plasser_detailed_2017,
etienne_theoretical_2017,
mai_quantitative_2018,
mewes_benchmarking_2018,
skomorowski_real_2018,
park_low_2018,
barca_excitation_2018,
etienne_comprehensive_2018}. In this context, a variety of quantities have been designed in order to unveil the nature of electronic transitions with scalars that provide -- through a simple number -- particular information related to the light-induced electronic structure reorganization \cite{luzanov_interpretation_1980,
peach_excitation_2008,
luzanov_electron_2009,
le_bahers_qualitative_2011,
garcia_evaluating_2013,
guido_metric_2013,
etienne_toward_2014,
etienne_new_2014,
guido_effective_2014,
etienne_probing_2015,
plasser_statistical_2015,
wenzel_physical_2016,
poidevin_truncated_2016,
plasser_entanglement_2016,
savarese_metrics_2017,
pastore_unveiling_2017,
etienne_theoretical_2017,
plasser_detailed_2017,
mai_quantitative_2018,
barca_excitation_2018,
mewes_benchmarking_2018,
etienne_charge_2018,
campetella_quantifying_2019}. An example of information that one might wish to seek when considering a molecular system undergoing a photoinduced intramolecular charge transfer is the locality of this charge transfer, i.e., the extent to which the electronic cloud has been perturbed and polarized \cite{etienne_toward_2014,etienne_new_2014,etienne_probing_2015,etienne_theoretical_2017,etienne_charge_2018}. To such probing of the charge transfer locality one can for example join the extent to which the detached and attached charges contribute to the net charge displacement occurring during an electronic transition \cite{etienne_probing_2015,etienne_theoretical_2017,etienne_charge_2018}. These two pieces of information are contained in two separate quantities, named $\phi _S$ and $\varphi$, respectively. \\ 
\indent Whilst the possibility of using computational resources for reaching such theoretical insights has already been used in theoretical and applied research, three different implementations were mentioned for evaluating these two density-based descriptors \cite{etienne_probing_2015}. Two of them were based on the population analysis \cite{mulliken_electronic_1955,
mulliken_electronic_1955-1,
mulliken_criteria_1962,
davidson_electronic_1967,
reed_natural_1985,
huzinaga_extended_1990,
parrondo_natural_1994,
clark_density_2004,
mayer_lowdin_2004,
bruhn_lowdin_2006,
bachrach_population_2007,
saha_are_2009,
jacquemin_what_2012} of the so-called detachment and attachment one-body reduced density matrices \cite{head-gordon_analysis_1995}, and the third one was the direct numerical integration of the related one-body charge density functions using Cartesian grids. When first published \cite{etienne_probing_2015}, the population analysis approximations to our density-based descriptors were compared to the numerical integration method as a reference solely for organic push-pull molecules using a single level of theory, for the sake of demonstrating its hypothetical usefulness, given that its implementation allows an extremely fast evaluation of the $\phi _S$ and $\varphi$ descriptors. \\
\indent In this contribution we wish to extend this diagnosis to a much wider set made of various types of molecules, and bring a rigorous and extensive diagnosis of the population analysis as an evaluation tool for our density-based descriptors. For this sake, we will use and combine in this paper multiple diagnosis criteria, and provide a precise motivation for the choice of the population analysis to perform, as well as a discussion based on the understanding of the possible preliminary transformations that are applied to the basis used before the theoretical evaluation of the descriptors is performed using the population analysis we selected. These principles will be involved in the critical discussion of the reliability of this evaluation path, which might be discarded in certain situations. With this we hope to provide the reader and user with the possibility to rationalize one's choice among the different possible derivations of the descriptors. \\
\indent This report is organized as follows: after bringing a short reminder on density matrices and density functions, as well as the possibility to rewrite a density matrix in Löwdin's symmetrically-orthogonalized basis, we will bring few mathematical tools that will be useful for writing the remaining expressions in the text in the most compact way possible. This technical part will also be the opportunity to justify certain choices for writing the descriptors in previous contributions. Afterward, the $\phi _S$ and $\varphi$ descriptors (among others) will be introduced, together with their population-analysis approximate value. Then, a general comment about the choice of the rules to diagnose in this contribution will be given, before the detailed diagnosis strategy is fully exposed in a dedicated section. Subsequently, the results of this diagnosis are introduced and discussed, before we conclude with the recommendations one could make to the future user of the related descriptors, with regards to the choice of the relevant evaluation method to use according to the system of interest. With this in hand, and taking into account certain restrictions related to the systems themselves, one should be able to consider the population-analysis approximation as being self-diagnosed — the population-analysis value of the descriptors is a partial measure of the reliability of the population-analysis approximation itself.

\section{Theoretical introduction}

\noindent In this contribution, we will bring few modifications to the nomenclature used until now in Refs. \cite{etienne_toward_2014,etienne_new_2014,etienne_probing_2015,pastore_unveiling_2017,etienne_theoretical_2017,
etienne_charge_2018}. These amendments are fully detailed in Appendix \ref{sub:appNom} with explanations about the reason why they were used then, and why they will no longer be employed.

\subsection{Brief reminder on density matrices and density functions}

\noindent 
In a finite, orthonormal basis $\left( \chi _r \right) _{r \in \llbracket 1, L\rrbracket}$ of one-particle spinorbital spatial parts, from a one-particle reduced density matrix (1--RDM \cite{cioslowski_many-electron_2000,
coleman_structure_1963,
davidson_reduced_1976,
lowdin_quantum_1955,
mcweeny_recent_1960,
mcweeny_methods_1992,
mcweeny_density_1959,
mcweeny_density_1961}) $\bm{\gamma}$, one can unequivocally give the expression of the corresponding one-body charge density (1--CD) function  $n$ defined on $\mathbb{R}^3$
\begin{equation}
\textbf{r} \longmapsto n(\textbf{r}) = \sum _{r=1}^L\sum _{s=1}^L (\bm{\gamma})_{rs} \, \chi _r (\textbf{r}) \chi _s^*(\textbf{r}).
\end{equation}
If the spatial part of every spinorbital function is expanded in the normalized but non-orthogonal basis of atomic functions $\left( \phi _r \right) _{r \in \llbracket 1,  K\rrbracket}$ as
\begin{equation}
\chi _r (\textbf{r}) = \sum _{\mu=1}^K (\textbf{C})_{\mu r} \, \phi _\mu (\textbf{r}) \qquad (1 \leq r \leq L)
\end{equation}
where the $r^\mathrm{th}$ column of \textbf{C} stores the so-called LCAO (standing for Linear Combination of Atomic Orbitals) coefficients of the $r^\mathrm{th}$ spinorbital, the 1--CD function can be recast as
\begin{equation}
n(\textbf{r}) = \sum _{\mu = 1}^K\sum _{\nu = 1}^K (\textbf{P})_{\mu \nu} \, \phi _\mu (\textbf{r}) \phi _\nu^*(\textbf{r}) \label{eq:nphi}
\end{equation}
with
\begin{equation}
\textbf{P} = \textbf{C}\bm{\gamma}\textbf{C}^\dag. \label{eq:Pgamma}
\end{equation}
The so-called overlap matrix \textbf{S} stores the spatial overlap between pairs of atomic functions, and its knowledge allows the straightforward evaluation of the integral of the 1--CD function over the whole space: Writing
\begin{equation}
\forall (\mu , \nu) \in \llbracket 1, K \rrbracket^2 , \,(\textbf{S})_{\mu \nu} := \int _{\mathbb{R}^3}  d\textbf{r} \; \phi _\mu ^* (\textbf{r}) \phi _\nu ^{\textcolor{white}{*}}  (\textbf{r})  
\end{equation}
implies
\begin{align} \nonumber
\int _{\mathbb{R}^3} d\textbf{r}\, n(\textbf{r}) = \mathrm{tr}\left(\textbf{PS}\right). \label{eq:trPS}
\end{align}
For any real number $x$, we have $\textbf{S} = \textbf{S}^{1-x}\textbf{S}^x$ and, according to the invariance of the trace of a matrix product upon cyclic permutations,
\begin{equation}
 \int _{\mathbb{R}^3}  d\textbf{r} \; n(\textbf{r}) = \mathrm{tr} \left( \textbf{S}^x \textbf{P}\textbf{S}^{1-x}\right) . \label{eq:SxPS1-x}
\end{equation}
In particular for ($x=1/2$), we have that $\textbf{S}^{1/2}\textbf{P}\textbf{S}^{1/2}$ corresponds to the 1--RDM in the so-called Löwdin symmetrically-orthogonalized basis of atomic orbitals \cite{lowdin_nonorthogonality_1950,
kashiwagi_generalization_1973,
scofield_note_1973,
scharfenberg_new_1977,
aiken_lowdin_1980,
attila_szabo_modern_1996,
mayer_lowdins_2002,
mayer_simple_2003,
mayer_lowdin_2004,
bruhn_lowdin_2006,
szczepanik_several_2013}.

\subsection{Tools for writing the density-based descriptors}

\noindent In order to simplify the notations in this contribution, we will define two functionals, that will be used throughout this paper. \\
\indent The first functional is constructed by considering a real-valued function $f$ defined on $\mathbb{R}^3$. Its value in every point of space is used for constructing two others, $f_-$ and $f_+$, according to the sign of the $f$ entries:
\begin{equation}
\textbf{r} \longmapsto  f_\pm (\textbf{r}) =   \left| f(\textbf{r}) \right| \pm f(\textbf{r}) .
\end{equation}
 Note that, for every $\textbf{r}$ in $\mathbb{R}^3$,
\begin{equation}
f_+(\textbf{r})-f_-(\textbf{r}) = f(\textbf{r}) ; \; f_+(\textbf{r}) + f_-(\textbf{r}) = \left| f(\textbf{r}) \right|.
\end{equation}
Averaging the integrals of $f_+ $ and $f_- $ over $\mathbb{R}^3$ gives the $G[f]$ functional
\begin{align}
G [f] &:= \dfrac{1}{2} \left(\int _{\mathbb{R}^3} d\textbf{r} \; f_{  +}(\textbf{r}) +  \int _{\mathbb{R}^3} d\textbf{r} \; f_{  -}(\textbf{r})\right) \nonumber\\&\textcolor{white}{:}=
\dfrac{1}{4} \int _{\mathbb{R}^3} d\textbf{r} \!\! \sum _{{\tau \in\left\lbrace -1,1\right\rbrace}} \underbrace{ \left| f (\textbf{r})\right| + \tau f (\textbf{r})}_{\displaystyle{2 f}_{\!\displaystyle \tau}(\textbf{r})}  . \label{eq:G}
\end{align}
This is basically how four of our density-based descriptors were defined in Refs. \cite{etienne_new_2014,etienne_probing_2015,pastore_unveiling_2017} (their relation to those reported here is discussed in Appendix \ref{sub:appGH}). We can see however that $G[f]$ can be reduced to a simple integral, according to 
\begin{equation}
\sum _{{\tau \in\left\lbrace -1,1\right\rbrace}} \tau \int _{\mathbb{R}^3} d\textbf{r} \,  f(\textbf{r}) = 0 \; \Longleftrightarrow \;  G[f] = \dfrac{1}{2} \int _{\mathbb{R}^3} d\textbf{r} \, \left|f(\textbf{r}) \right| .
\end{equation}
The reason why those descriptors were computed as in Eq. \eqref{eq:G} is that, although the functions considered were satisfying
\begin{equation}
\int _{\mathbb{R}^3} d\textbf{r} \,  f(\textbf{r}) = 0 \; \Longleftrightarrow \; \int _{\mathbb{R}^3} d\textbf{r} \,  f_+(\textbf{r}) = \int _{\mathbb{R}^3} d\textbf{r} \,  f_-(\textbf{r}),
\end{equation}
in practice the integrals of $f_+$ and $f_-$ were averaged together in order to decrease the numerical imprecision inherent to their practical evaluation.\\
\indent For the second functional used in this report, we will consider any square, real matrix $\textbf{Q} \, \in \, \mathbb{R}^{m \times m}$, and define 
\begin{equation}
H [\textbf{Q}] := \dfrac{1}{4} \sum _{{\tau \in\left\lbrace -1,1\right\rbrace}} \!\! \mathrm{tr} \left[ \sqrt{\left(\textbf{Q}\circ I_m\right)^2} + \tau \left(\textbf{Q} \circ I_m\right) \right] 
\end{equation}
where this time the integration has simply been replaced by a trace. The ``$\circ${\textquotedblright} symbol stands for the Hadamard product (sometimes called the entrywise product), defined as
\begin{equation}
\left(\textbf{Q}\circ \textbf{E}\right)_{ij} = (\textbf{Q})_{ij}(\textbf{E})_{ij},
\end{equation}
and $I_m$ is the $m \times m$ identity matrix, so that $\textbf{Q}\circ I_m$ is a diagonal matrix whose nonzero elements are the diagonal elements of $\textbf{Q}$ only, and
\begin{equation}
\left(\textbf{Q}\circ I_m\right)_{ij} =   \left(\textbf{Q}\right)_{ij} \delta _{ij}, 
\left( \sqrt{\left(\textbf{Q}\circ I_m\right)^2}\,\right)_{\!\!ij} = \left| \left(\textbf{Q}\right)_{ij} \right| \delta _{ij},
\end{equation}
so that in practice, $H[\textbf{Q}]$ simply isolates the diagonal of \textbf{Q} in order to average its positive and negative diagonal entries sums. For the same reason as for $G$ this procedure was used in Ref. \cite{etienne_probing_2015} to compute most of the matrix-algebraic approximations to the descriptors (their link to those derived here is highlighted in Appendix \ref{sub:appGH}) though, again, we see that we can simplify its expression:
\begin{equation}
\sum _{{\tau \in\left\lbrace -1,1\right\rbrace}} \!\! \tau\, \mathrm{tr}\left(\textbf{Q} \circ I_m\right)  = 0 \; \Longleftrightarrow \;  2H[\textbf{Q}] =  \mathrm{tr}\sqrt{\left(\textbf{Q}\circ I_m\right)^2}. \label{eq:H}
\end{equation}
In this contribution, $G$ and $H$ will be used to revisit and generalize various descriptors of electronic transitions and to express their derivation using two types of procedures: numerical integration on a Cartesian grid ($G$) and matrix-algebraic approximation ($H$).\\
\indent Finally, note that the $G$ and $H$ functionals are similar mappings in a continuous and discrete basis, respectively.

\subsection{The target density-based descriptors}

\noindent We start by considering the so-called detachment and attachment 1--RDMs. Those can be seen as the hole and particle density matrices, in a one-hole/one-particle representation of an electronic transition. They are constructed by diagonalizing the one-body difference density matrix (1--DDM, $\bm{\gamma}^\Delta$), i.e., the matrix obtained, in the spinorbital space, by subtracting the ground electronic state 1--RDM ($\bm{\gamma}^0$) to the one corresponding to the $q^\mathrm{th}$ excited state ($\bm{\gamma} ^q$)
\begin{equation}\label{eq:gammaDelta}
\textcolor{black}{\bm{\gamma} ^\Delta} = \bm{\gamma} ^q - \bm{\gamma} ^0 \; \in \; \mathbb{R}^{L \times L} ,
\end{equation}
\begin{equation} \exists (\textbf{M},\textbf{K}) \in \left(\mathbb{R}^{L\times L}\right)^2 ,\, \textbf{M}^\dag \textcolor{black}{\bm{\gamma} ^{\Delta}} \textbf{M}  = \mathrm{diag}(K_i)_{i\in\llbracket 1, L \rrbracket}.
\end{equation} 
where the columns of $\textbf{M}$ are called natural difference orbitals. This operation also produces eigenvalues, which are often referred to as transition occupation numbers, and can be sorted with respect to their sign:
\begin{equation}
\textbf{K}_\pm = \dfrac{1}{2} \left(\sqrt{\textbf{K}^2} \pm \textbf{K}\right).
\end{equation}
This operation is similar to those involved in the construction of $G$ and $H$. Backtransformation provides the $L\times L$ real-valued {detachment} (\textcolor{black}{$\bm{\gamma} ^d$}) and {attachment} (\textcolor{black}{$\bm{\gamma} ^a$}) 1--RDMs
\begin{equation}
\textbf{M}\textcolor{black}{\textbf{K}_-}\textbf{M}^\dag = \textcolor{black}{\bm{\gamma} ^d}; \; \textbf{M}\textcolor{black}{\textbf{K}_+}\textbf{M}^\dag = \textcolor{black}{\bm{\gamma} ^a} . \label{eq:gamma_da}
\end{equation} 
Once projected in the Euclidean space, they give the detachment and attachment 1--CD functions:
\begin{align}\nonumber
 n_d (\textbf{r}) &= \sum _{r=1}^L\sum _{s=1}^L (\bm{\gamma}^d)_{rs} \, \chi _r (\textbf{r}) \chi _s ^* (\textbf{r}) \\ &= \sum _{\mu =1}^K\sum _{\nu =1}^K (\underbrace{\textbf{C}\bm{\gamma}^d \textbf{C}^\dag}_{\displaystyle{\textbf{D}}})_{\mu \nu} \, \phi _\mu (\textbf{r})\phi _\nu ^* (\textbf{r}) \label{eq:nd}
\end{align}
and
\begin{align} \nonumber
 n_a (\textbf{r}) &= \sum _{r=1}^L\sum _{s=1}^L (\bm{\gamma}^a)_{rs} \, \chi _r (\textbf{r}) \chi _s ^* (\textbf{r}) \\ &= \sum _{\mu =1}^K\sum _{\nu =1}^K (\underbrace{\textbf{C}\bm{\gamma}^a \textbf{C}^\dag}_{\displaystyle{\textbf{A}}})_{\mu \nu} \, \phi _\mu (\textbf{r})\phi _\nu ^* (\textbf{r}). \label{eq:na}
\end{align}
which are both positive or zero everywhere in $\mathbb{R}^3$. We have, for every real number $x$,
\begin{align}  \nonumber
\vartheta &= \int _{\mathbb{R}^3} \hspace*{-0.2cm} d\textbf{r} \; n_d (\textbf{r}) = \mathrm{tr} \left(\textbf{S}^x \textbf{D}\textbf{S}^{1-x}\right) \\ &= \int _{\mathbb{R}^3} \hspace*{-0.2cm} d\textbf{r} \; n_a (\textbf{r}) = \mathrm{tr} \left(\textbf{S}^x \textbf{A}\textbf{S}^{1-x}\right)  \label{eq:vartheta}
\end{align}
This $\vartheta$ represents the quantity of charge that has been involved in the electronic transition. It is this integral of the detachment and attachment densities that will be used as a normalization factor for the quantities defined below. Note that in previous publications \cite{etienne_toward_2014,etienne_new_2014,
etienne_probing_2015,etienne_theoretical_2017,
pastore_unveiling_2017,etienne_charge_2018} the integral of the detachment and attachment densities was averaged to give $\vartheta$, for the same reasons as those described in the penultimate paragraph of Appendix \ref{sub:appNom}.\\
\indent Note that here ``\textbf{A}{\textquotedblright} and ``\textbf{D}{\textquotedblright} should not be confused with electron ``Acceptor{\textquotedblright} and ``Donor{\textquotedblright}. \\
\indent We can now write two functions involving the detachment/attachment densities
\begin{equation}
n_{\Delta} (\textbf{r}) = n_{a}(\textbf{r}) - n_{d}(\textbf{r}) \; \in \, \mathbb{R} \; ; \; \eta  (\textbf{r}) =  \sqrt{n_{d}(\textbf{r})n_{a}(\textbf{r})}  \; \in \, \mathbb{R}^+, \label{eq:nDelta}
\end{equation}
where
\begin{equation}
n_{\Delta} (\textbf{r}) = n_{n}(\textbf{r}) - n_{0}(\textbf{r}),
\end{equation}
and write the detachment/attachment overlap integral \cite{etienne_toward_2014}
\begin{equation}
\phi _S   := \vartheta ^{-1}\int _{\mathbb{R}^3} d\textbf{r} \; \sqrt{n_{d}(\textbf{r})n_{a}(\textbf{r})} = 2 \vartheta ^{-1} G[\eta] \; \in \; [0,\!1]
\end{equation}
as well as the non-normalized \cite{etienne_new_2014} ($\chi$) and normalized \cite{etienne_probing_2015} ($\varphi$) detachment/attachment-based charge-displacement descriptor
\begin{align}
{\chi} := G[n_\Delta] \; ; \;  \vartheta   \geq {\chi}  \; \Longrightarrow \; \exists ! \varphi \in \; [0,\!1], \, {\varphi}  = \vartheta^{-1} {\chi}.
\end{align}
$\varphi$ measures the fraction of detachment/attachment density contributing to the net transferred charge. $\phi _S $ and $\varphi$ were then combined in Ref. \cite{etienne_probing_2015} into the normalized, general $\psi  $ descriptor
\begin{equation}
\psi   := \left(\pi/2\right) ^{-1} \mathrm{arctan}\left(\dfrac{\phi _S  }{{\varphi} }\right) \; \in \; [0,\!1[
\end{equation}
jointly describing the range of a charge separation, and the amount of charge transferred during an electronic transition.

\subsection{Alternative approach: the population analysis}

\noindent Since it is not possible to solve analytically the $\phi _S$ and $\varphi$ integrals reported above, only approximate values can be obtained. In this paper we extend the matrix-algebraic procedure proposed in Ref. \cite{etienne_probing_2015}, where numerical integrations on Cartesian grids usually used for computing the density-based descriptors $\phi_S$ and $\varphi$ were approximated by performing a detachment/attachment population analysis (these two methods were loosely called direct- and Hilbert-space derivations in Ref. \cite{etienne_probing_2015}, the latter being employed because one-body orbital wave function integrals were part of the derivation procedure): If the atomic functions are centered on atoms, a known procedure is to approximate the atomic population, i.e., the partial charge that an atom bears for a given electronic distribution, through a Mulliken population analysis by multiplying a density matrix with the basis set overlap matrix \textbf{S}, and to consider the $k^\mathrm{th}$ diagonal entry of this product of matrices as the contribution of the $k^\mathrm{th}$ atomic orbital to the electronic population of the system. The actual approximation then consisted in considering the fact that the basis of atomic functions was local, with the Gaussian basis functions grossly covering a given region of space, so that manipulating (in particular, subtracting or multiplying) the diagonal entries of the matrix product of the atomic-space detachment and attachment density matrices with \textbf{S} was an approximation to the manipulation of elementary fractions of a charge density in a given region of space. Another possible approximation reported in Ref. \cite{etienne_probing_2015} is the Löwdin detachment/attachment population analysis, where the density matrices are contracted to the left and to the right by the square root of \textbf{S} to orthogonalize the basis set while keeping as much as possible its local character. In fact, any combination of real-valued exponents for the left and right multiplication can arbitrarily be used, as long as the sum of the exponents is equal to unity, in order to satisfy Eq. \eqref{eq:SxPS1-x}. \\
\indent Though the choice of the $x$ exponent seems quite arbitrary, it was demonstrated in Appendix \ref{sub:appLow} that only a Löwdin-like population analysis produces atomic populations that are not susceptible of being negative or greater than the maximum allowed occupancy, which would be unphysical. Therefore, though the matrix-algebraic definition of the approximate descriptors are given below for any value of $x$, in the next section of this paper we will solely use the Löwdin-like scheme (denoted by the ``$\ell${\textquotedblright} symbol), unless the contrary is explicitly stated.\\
\indent  The limitations of the approximation made by using symmetrically-orthogonalized orbitals populations as if we were mapping these charges to point charges in space when computing our descriptors is diagnosed and discussed in this report.\\
\indent According to Eq. \eqref{eq:SxPS1-x}, since the integral of a 1--CD function is unequivocally equal to the trace of the product of \textbf{S} with the density matrix corresponding to the charge distribution when this matrix is known, the $\vartheta$ index is computed without approximation as the trace of the unrelaxed difference density matrix multiplied by \textbf{S}, using $x=0$ for sparing the effort of diagonalizing \textbf{S}. \\
\indent The population analysis approximations to the detachment/attachment spatial overlap $\phi _S$ are 
\begin{equation}
\phi_S^{(x)} = \vartheta ^{-1} \, \mathrm{tr} \sqrt{\left( \bm{\Omega}_x \circ I_K \right)} 
\end{equation}
with $x \in \mathbb{R}$, and $\bm{\Omega}\in \mathbb{R}^{K\times K}$, with, for every $(\mu , \nu)$ in $\llbracket 1,K \rrbracket ^2$,
\begin{equation}\label{eq:phiSx}
\left(\bm{\Omega}_x \right)_{\mu \nu} = \prod _{\bm{\nu} \in \left\lbrace \textbf{D},\textbf{A}\right\rbrace}(\textbf{S}^{x}\bm{\nu}\textbf{S}^{1-x})_{\mu \nu} \left[ (\textbf{S}^{x}\bm{\nu}\textbf{S}^{1-x})_{\mu \nu} > 0\right]   
\end{equation}
where the last factor is called ``Iverson's brackets{\textquotedblright}. This factor takes a value of zero when the condition between the brackets is not true, and a value of one when the condition is true. Note that the $\phi_S^x$ functional should not be confused with $H$ in Eq. \eqref{eq:H}. Note also that the Iverson brackets are imposed in Eq. \eqref{eq:phiSx} for every matrix element, though this condition is necessary only for the diagonal elements, because the non-diagonal entries are removed by the Hadamard product in the definition of the $H$ functional. This justification is transferrable to the derivation of each other quantity below.\\
\indent Due to the fact that solely the Löwdin analysis produces populations that cannot be negative, the Iverson brackets above are always equal to one for the diagonal elements of $\displaystyle{\textbf{S}^{1/2}\textbf{D}\textbf{S}^{1/2}}$ and $\displaystyle{\textbf{S}^{1/2}\textbf{A}\textbf{S}^{1/2}}$, so the Löwdin-derived approximation to $\phi _S$, written here $\phi _S^\ell$, reads
\begin{align} \nonumber
\phi _S ^\ell &= \vartheta ^{-1} \, \mathrm{tr}\sqrt{\left(\textbf{S}^{1/2}\textbf{D}\textbf{S}^{1/2} \circ I_K\right)   \left(  
\textbf{S}^{1/2}\textbf{A}\textbf{S}^{1/2} \circ I_K \right)} \\ &:= F[\textbf{D},\textbf{A},\textbf{S}].
\end{align}
If we are now interested in the computation of $\varphi$ with matrix algebra, we find that the $H$ functional defined in Eq. \eqref{eq:H} becomes very useful, since we can write the generalised (i.e., for any real $x$) approximation to $\varphi$ as
\begin{equation}
\varphi ^{(x)} = \vartheta ^{-1} \, \underbrace{H[\bm{\mu}_x]}_{\displaystyle \chi ^{(x)}} 
\end{equation}
with, for every $(\mu , \nu)$ in $\llbracket 1, K \rrbracket ^2$,
\begin{align}\nonumber
\left(\bm{\mu}_x\right)_{\mu \nu} &=  (\overbrace{\textbf{S}^{x} \textbf{A} \textbf{S}^{1-x} -  \textbf{S}^{x}\textbf{D}\textbf{S}^{1-x}}^{\displaystyle{\textbf{S}^x \bm{\Delta}\textbf{S}^{1-x}}} )_{\mu \nu} \\ &\times \left[\left(\textbf{S}^{x} \textbf{D} \textbf{S}^{1-x}\right)_{\mu \nu} > 0\right]    \left[\left(\textbf{S}^{x}\textbf{A} \textbf{S}^{1-x}\right)_{\mu \nu} > 0\right] 
\end{align}
where $\bm{\Delta}$ is the unrelaxed difference density matrix in the atomic space ($\textbf{C}\bm{\gamma}^\Delta \textbf{C}^\dag$). When the Löwdin analysis is concerned for approximating $\varphi$, we see that the two Iverson brackets are equal to unity, leading to
\begin{equation}
\varphi ^\ell = \vartheta ^{-1} \, \underbrace{H[\textbf{S}^{1/2}\bm{\Delta}\textbf{S}^{1/2}]}_{\displaystyle \chi ^\ell}.
\end{equation}

\subsection{Beyond the unrelaxed picture of electronic transitions}

\noindent If the description of the excited state electronic structure involves a pseudo-orbital relaxation through the addition of a Z-vector ($\bm{\gamma}^\mathrm{Z}$ in the spinorbital space) to the 1--DDM, one obtains the ``relaxed{\textquotedblright} difference, detachment, and attachment 1--RDMs. Those are obtained using the same procedure as in Eqs. \eqref{eq:gammaDelta} to \eqref{eq:gamma_da}. The corresponding 1--CDs $n' _\Delta$, $n'_d$ and $n'_a$ are also constructed using the same rules as in Eqs. \eqref{eq:nd}, \eqref{eq:na}, and \eqref{eq:nDelta}. Finally, the integral of the relaxed detachment/attachment 1--CDs, computed in a way similar to the one reported in Eq. \eqref{eq:vartheta}, will be written $\vartheta '$. \\
\indent In a recent contribution \cite{pastore_unveiling_2017} we were interested in describing the nature of this pseudo-orbital relaxation. For this purpose, we designed three quantities: the first one is the difference between the rearranged (detached/attached) charge in the relaxed and unrelaxed schemes,
\begin{equation}
 \lambda^\mathrm{Z} := \vartheta' - \vartheta.
\end{equation}
We also have looked at the relaxation-induced detachment/attachment rearranged charge  ($\bm{\nu} \in \left\lbrace \textbf{D},\textbf{A}\right\rbrace$)
\begin{equation}
\alpha ^\ddag _{\bm{\nu}} := G[n ^\ddag _{\bm{\nu}}] \; ; \; n ^\ddag _{\bm{\nu}} (\textbf{r}) = \lambda ^\mathrm{Z}\left[  \dfrac{n'_{\bm{\nu}}(\textbf{r})}{\vartheta '} - \dfrac{n_{\bm{\nu}}^{\textcolor{white}{e}}(\textbf{r})}{\vartheta}\right]  
\end{equation}
($\alpha^\ddag _{\bm{\Gamma}}$ and $\alpha^\ddag _{\bm{\Lambda}}$ in the original publication, see Appendix \ref{sub:appNom}), which corresponds to the separate impact of the Z-vector on the detachment and attachment densities. Finally, the third quantity used for unveiling the nature of the post-linear response pseudo-orbital relaxation, is the charge effectively displaced by the relaxation process:
\begin{equation}
\zeta ^\mathrm{Z} := G[n_\mathrm{Z}] 
\end{equation}
with
\begin{align}\nonumber
n_\mathrm{Z}(\textbf{r}) &= \sum _{r=1}^L\sum _{s=1}^L \left(\bm{\gamma}^\mathrm{Z}\right)_{rs} \, \chi _r (\textbf{r}) \chi _s ^* (\textbf{r})\\ &= \sum _{\mu =1}^K \sum _{\nu =1}^K (\overbrace{\textbf{C}\bm{\gamma}^\mathrm{Z} \textbf{C}^\dag }^{\displaystyle{\bm{\Delta}^\mathrm{Z}}})_{\mu \nu} \, \phi _\mu (\textbf{r})\phi _\nu ^* (\textbf{r}). \label{eq:zetaZ}
\end{align}
This $n_\mathrm{Z}$ density function actually corresponds to the projection of the Z-vector in the Euclidean space. \\
\indent If we now turn to the population analysis computation of these unnormalized descriptors we see that the $\zeta ^\mathrm{Z}$ quantity reported in Eq. \eqref{eq:zetaZ} can also be approximated using $H$. First, we rewrite the $\bm{\Delta}^\mathrm{Z}$ we met in Eq. \eqref{eq:zetaZ}
\begin{equation}
\bm{\Delta}^\mathrm{Z} = \bm{\Delta}' - \bm{\Delta} =  \left(\textbf{P}_q' - \textbf{P}_0 \right) -  \left(\textbf{P}_q - \textbf{P}_0 \right) = \textbf{P}_q' - \textbf{P}_q.
\end{equation}
where $\bm{\Delta}'$ is the atomic-space relaxed 1--DDM, $q$ is the excited state number (as in Eq. \eqref{eq:gammaDelta}), and each \textbf{P} is an atomic-space 1--RDM, related to the spinorbital-space 1--RDM by Eq. \eqref{eq:Pgamma}. With this, we find
\begin{equation}
\zeta ^\mathrm{Z}_x = H\left[\bm{\Xi}_x\right]
\end{equation}
with, for every $(\mu, \nu)$ in $\llbracket 1, K \rrbracket ^2$,
\begin{align}\nonumber
\left(\bm{\Xi}_x\right)_{\mu \nu} &=  (\textbf{S}^{x} \bm{\Delta}^{\mathrm{Z}} \textbf{S}^{1-x})_{\mu \nu} \\ &\times \left[\left(\textbf{S}^{x} \textbf{P}_q \textbf{S}^{1-x}\right)_{\mu \nu} > 0\right]    \left[\left(\textbf{S}^{x}\textbf{P}_q' \textbf{S}^{1-x}\right)_{\mu \nu} > 0\right] .
\end{align}
Our considerations about Löwdin' scheme for population analysis lead us to write
\begin{equation}
\zeta ^\mathrm{Z}_\ell = H\left[\textbf{S}^{1/2} \bm{\Delta}^{\mathrm{Z}} \textbf{S}^{1/2}\right].
\end{equation}
Finally, as far as the $\alpha ^\ddag _{\textbf{D}}$ and $\alpha ^\ddag _{\textbf{A}}$ descriptors are concerned, we give here the following approximation to their value:
\begin{equation}
\alpha ^\ddag _{\bm{\nu},x} = \lambda ^{\mathrm{Z}} H\left[ \bm{\varpi}^{\bm{\nu}}_x \right] 
\end{equation}
with, for every $(\mu, \nu)$ in $\llbracket 1, K \rrbracket ^2$,
\begin{align}\nonumber
\left(\bm{\varpi}^{\bm{\nu}}_x\right)_{\mu \nu} &= \left( \dfrac{\textbf{S}^x \bm{\nu}' \textbf{S}^{1-x}}{\vartheta '} - \dfrac{\textbf{S}^x \bm{\nu} \textbf{S}^{1-x}}{\vartheta} \right)_{\mu \nu}\\ &\times \left[\left(\textbf{S}^x \bm{\nu}' \textbf{S}^{1-x}\right)_{\mu \nu} > 0 \right] \left[\left(\textbf{S}^x \bm{\nu} \textbf{S}^{1-x}\right)_{\mu \nu} > 0 \right]
\end{align}
which is given for $\bm{\nu} \in \left\lbrace \textbf{D},\textbf{A}\right\rbrace$. The Löwdin scheme returns
\begin{equation}
\alpha ^\ddag _{\bm{\nu},\ell} =   H\left[ \bm{\varpi}^{\bm{\nu}}_\ell \right] \; \mathrm{with} \; \bm{\varpi}^{\bm{\nu}}_\ell  = \lambda ^\mathrm{Z} \left[ \dfrac{\textbf{S}^{1/2}\bm{\nu}'\textbf{S}^{1/2}}{\vartheta'} -  \dfrac{\textbf{S}^{1/2}\bm{\nu}\textbf{S}^{1/2}}{\vartheta}\right].
\end{equation}

\subsection{Diagnosis}

\noindent Each descriptor that has been given an approximation with population analysis above is either derivable exactly ($\vartheta$, $\vartheta '$, and $\lambda ^\mathrm{Z}$), or by using the $H$ rule ($\varphi ^\ell$, $\zeta ^\mathrm{Z}_\ell$, $\alpha ^\ddag _{\bm{\nu},\ell}$) or the $F$ one ($\phi _S^\ell$). Therefore, the diagnosis performed for this report actually concerns the $H$ and $F$ rules. In particular, for the $H$ rule we selected to investigate the $\varphi ^\ell$ approximation, since the $\varphi ^\ell$ number can be combined to $\phi _S^\ell$ to produce $\psi ^\ell$ without performing any additional matrix manipulation. It is assumed here that since the evaluation of $\zeta ^\mathrm{Z}_\ell$ and the two $\alpha ^\ddag _{\bm{\nu},\ell}$ quantities follows computational schemes identical to the one of $\varphi ^\ell$, our diagnosis on the latter also holds for the former. \\
\indent Finally, note that due to Iverson's brackets in the expression of the matrix elements of $\bm{\Omega}_x$, $\bm{\mu}_x$, $\bm{\Xi}_x$, and $\bm{\varpi}_x$, negative populations are ignored in the computation of the related descriptors, so that there might be occurrences in which the trace of the matrices on the elements of which Iverson's brackets build conditions is not preserved since these negative populations can be overbalanced in other (positive) diagonal entries that are not ignored by Iverson's brackets. This also means that these negative entries, which actually contribute to the value of the relevant 1--CD in every point of space, are conveniently ignored as they would appear below a square root in $H$. Fortunately, such case never occurs with the Löwdin scheme we use in practice and in this paper (see Appendix \ref{sub:appLow}).

\section{Diagnosis strategy}

\noindent Our diagnosis strategy is summarized in Table \ref{tableau 1}. All the calculations reported in this paper were performed in vacuum, using the Gaussian16 (revision A03) package \cite{g16}. A set of fifty-nine metal-free molecules of varying size (from diatomics to complex molecular systems) was used in order to perform our diagnosis. The Cartesian grids were generated using the Cubegen utility from Gaussian16, and the electronic-structure analyses (numerical integration - NI - of the Cartesian grids, and population analyses - PA) were performed using a homemade code, to be released soon.\\
\indent In order to determine to which extent both methods are consistent, different variables have been chosen for comparing NI and PA values: the system size, the charge transfer (CT) nature, the basis set (BS) size and nature, and the type of exchange-correlation (xc) functional used in time-dependent density functional theory. \\
\indent Through this study, the influence of the Cartesian grid density of points has been evaluated by comparing medium (6 points/bohr for each direction) grids to fine (12 points/bohr for each direction) ones. \\
\indent Our first molecular test-set ({S1}) is taken directly from Ref. \cite{etienne_toward_2014} and is constituted by a series of conjugated molecules with nitro (-NO$_2$) acceptor and dimethylamino (-NMe$_2$) donor terminal groups, separated by a varying number (ranging from one to five) of bridge subunits. The level of theory used here for our computations is exactly the one from the reference cited above. \\
\indent One of the spacers is an oligo-ethylene for the {nIII} group of molecules. The oligo-ethylene might separate two donor groups for what we will call the nIII-{NMe$_2$-NMe$_2$} series here, or two nitro groups (nIII-NO$_2$-NO$_2$), or one of each (nIII-NMe$_2$-NO$_2$). \\
\indent If the donor and the acceptor are separated by a varying number of phenyl moieties, we have the nVI-c series. \\ 
\indent Finally, in the nV-X series, the dimethylamino and nitro groups are connected through an heterocyclic oligomer (with five-member rings containing one oxygen, sulfur, or selenium) or through oligo-pyrrols. \\
\indent The equilibrium geometries of the molecules belonging to the S1 set were obtained using the PBE0/6-311G(d,p) level of theory. Vibrational frequencies have been calculated to verify that the optimized 
geometries are actual minima. The excited-state calculations were done at the PBE0/6-311++G(2d,p) level of theory. \\
\indent The particular interest of this set lies in the possibility to investigate electronic transitions of different nature through a series, with molecules of different size, hence calculations with basis sets of different size, so that the only stable variable is the xc-functional through the diagnosis. When switching the heteroatom in the nV-X series from oxygen to sulfur or selenium, one can also see how an increase of the number of electrons and basis functions for molecules of comparable size influences the precision of the PA approximation. This will be our first diagnosis, denoted $d_1$ in the results section, as well as in Table \ref{tableau 1} below. \\
\indent In addition to this, excited-state calculations were performed on the nV-X series using a small- (6-31G) and intermediate-sized (6-31+G(d,p)) BS in order to check the BS convergence for the PA values at a given geometry for this type of push-pull molecule ($d_2$). \\
\indent The second test-set (S2) originates from Ref. \cite{peach_excitation_2008} and will be investigated using the same level theory as in the original paper, unless the contrary is explicitly stated. In particular, we will investigate the diatomic (N$_2$, CO, and HCl) and formaldehyde molecules from S2 in order to diagnose the reliability of $F$ and $H$ rules for very small molecules ($d_3$). From S2 we will also focus on the acenes ($n+1$ fused benzene rings, with $n$ ranging from one to five) series, as we know that for such molecules the detachment and attachment densities will exhibit a substantial spatial overlap, no matter the number of subunits. This diagnosis will allow us to know whether, for such an extreme case of transition nature, there might be an influence from the size of the system on the accuracy of the PA approximation ($d_4$). Note that we also performed excited-state calculations on the acenes series using the B3LYP/6-31G and B3LYP/6-311++G(2d,p) levels of theory in order to assess the basis set convergence on the reference and approached descriptor values ($d_5$).\\
\noindent The last system that was used for this report is the pyridinium phenolate (PP) that was already investigated in Ref. \cite{pastore_unveiling_2017}. We used a rigid scan of the central dihedral angle, scanning the angle from zero to ninety degrees with a one-degree step, in order to see the evolution of the nature of the donor-acceptor intramolecular charge transfer. Such an approach allowed us to investigate a system of fixed size, at a given level of theory, with tunable CT character, for a one-variable diagnosis (\textit{d$_6$}) on a medium-sized molecular system. This study was also repeated including a variation of the level of theory by performing this ninety-one-point scan for a combination of fifteen split-valence Pople basis sets \cite{ditchfield_selfconsistent_1971,frisch_selfconsistent_1984} and six xc-functionals (the hybrid B3LYP \cite{becke_densityfunctional_1993}, PBE0 \cite{adamo_toward_1999}, and M06-2X \cite{zhao_m06_2008}, the long-range corrected CAM-B3LYP \cite{yanai_new_2004} and LC-$\omega$HPBE \cite{vydrov_importance_2006}, and the range-separated, dispersion-corrected $\omega$B97X-D \cite{chai_long-range_2008}), so that we could select some geometries and evaluate the separate influence of the basis set size (and nature), as well as the type (hybrid/range-separated hybrid) of xc-functional (\textit{d}$_7$ and $d_8$). Configuration Interaction with single excitations (CIS \cite{hirata_configuration_1999}) and Time-Dependent Hartree-Fock (TDHF \cite{hirata_configuration_1999}) calculations were also performed on PP, combined with the fifteen BSs used in TDDFT \cite{hirata_configuration_1999} above. \\
\indent The sketch of the molecules investigated in this report is given in \textcolor{black}{figure S1}.

\section{Results}

\noindent Due to our practical use of the descriptors, we report herafter their values rounded to two decimal places and, when comparing two evaluation paths, we will consider the consistency between the two methods as achieved when the results are consistent within two decimal places. Moreover, consistently with the discussion initiated in Ref. \cite{etienne_probing_2015}, we will consider a tolerance for the absolute deviation of $\pm$ 0.05 arbitrary units for the descriptors as the maximum allowed deviation between two methods.

\subsection{Influence of the NI grid density of points}

\noindent When studying the whole set of molecular systems for this report, we quasi-systematically used the two density of points (6 and 12 points/bohr in each direction) for the sake of comparison and, according to our convergence/consistency criterion enonciated above, we noticed that the results on the descriptors are strictly converged for medium-density grids (6 points/bohr in each direction) and can be considered exact for the level of theory used, which means that the user who wishes to compute our descriptors can reliably spare the cost of producing fine grids eight times denser than the medium-density one. We drag the attention of the reader on the fact that when the numerical integration is performed, one has always to carefully check whether the integral of the detachment and attachment densities are strictly equal. We report in SI (see \textcolor{black}{Tables S29 to S34}) our data for the nIII group of molecules, and show that for the nIII-NMe$_2$-NMe$_2$ series and one excited state of the nIII-NMe$_2$-NO$_2$ series the standard grid sizes lead to a significant fraction of charge lost in the attachment. The integrals obtained with finer grids are also provided in order to prove that the fraction of charge lost comes from the grid size instead of the grid density of points. In order to overcome such issue the user should increase the grid size with the same density of points to get the right detachment/attachment integral. In this regard one can use the exactness of the trace of \textbf{DS} and \textbf{AS} to diagnose the quality of the integration grid, as it was done in \textcolor{black}{Tables S29, S31, and S33}. 

\subsection{The $F$ and $H$ rules}

\noindent Again, when considering the total set of data that we report in Supplementary Information for the fifty-nine molecules studied in this paper, we observed that, according to our accuracy/consistency criterion mentioned at the beginning of this section, when the PA-derived $\phi_S$ values were found to be reliable, so were the $\varphi$ values, though they were not obtained from the exact same rule. Therefore, from now on in this section we solely will discuss the $\phi _S$ and $\phi _S^\ell$ values for the sake of brevity, though the data in the tables and in the Supplementary Information will also contain the $\varphi$ and $\psi$ diagnosis.  

\subsection{The oligo-acenes series}

\noindent As expected, when studying the oligo-acenes series with our descriptors, we noticed that for the three xc-functionals used, i.e., PBE, B3LYP and CAM-B3LYP, in the $d_4$ diagnosis (see Table \ref{tableau 1}), the electronic cloud remains almost unperturbed by the electronic transitions, leading to $\phi _S$ values systematically greater than 0.9, which is quite an extreme case of detachment/attachment spatial overlap. This statement holds irrespective of the size of the molecule and of the BS (diagnosis $d_5$), or of the xc-functional used. Results in Table \ref{tableau 2} and in \textcolor{black}{Tables S1 to S5} show that in such an extreme case the reliability of the PA to approximate the value of $\phi _S$ cannot be firmly established. Indeed, we saw among the results both exact NI/PA matching and deviations beyond our tolerance criterion, which means that for such extreme situations the use of the PA-approximation $\phi _S^\ell$ is strongly discouraged. These conclusions also hold for the poly-acetylene series from the same S2 set of molecules (\textcolor{black}{see Tables S38 to S40}).

\subsection{Influence of the size of the system}

\noindent When the nV-X and nVI-c sets of molecular systems are concerned, we globally observe that the molecular composition influences the nature of the CT, and that when performing the $d_1$ diagnosis the $\phi _S ^\ell$ approximation can become (but is not systematically) problematic both for small systems and for electronic transitions characterized by a small $\phi _S ^\ell$. \\
\indent For the nV-X and nVI-c series (see Table \ref{tableau 3}, Tables S27 and S28, S35 to S37, and S44 to S47) we observe that for medium- to large-sized systems ($n > 3$) the $\phi _S^\ell$ descriptor reproduces very accurately the $\phi _S$ value. Conversely, for small systems ($n = 1$ or $n=2$) the PA-approximation is no longer reliable for $\phi _S$ values lower or equal to 0.52. This was also observed for the oligo-peptides of the S2 set of molecules (\textcolor{black}{see Tables S6 to S8, S12 to S14, and S48 to S50}). Note that in the S2 set of molecules the geometrical origin of the NI-PA deviation also applies for higher $\phi _S$ values, for example when the CO, N$_2$, HCl, and H$_2$CO molecules (see Tables S9 to S11, and S18 to S26), which are too small for considering the $\phi _S^\ell$ approximation as reliable, are concerned (diagnosis $d_3$).\\
\indent The $d_2$ diagnosis (\textcolor{black}{see Tables S27 and S28, S36 and S37, and S44 to S47}) allowed us to conclude that the BS convergence was rapidly achieved for the $\phi _S$ value with medium-sized BS, and that when comparing the converged $\phi _S$ values to their PA-approximated values, the accuracy of the PA approximation was independent from the BS size. In other words, if the level of theory contains a BS of intermediate or large size, increasing the size of the BS did not improve the accuracy of the PA, \textcolor{black}{as it was already observed in the $d_5$ diagnosis.} Following the same idea, we notice in Table \ref{tableau 3} that replacing oxygens by sulfur or selenium might alter the $\phi _S$ value but not the accuracy of its PA-approximation, although in such substitutions we considerably modify the number of basis functions centered on the heteroatom positions before the symmetric orthogonalization.

\subsection{Study of medium-sized molecules}

\noindent The accurate $\phi _S ^\ell$ results for the S2-belonging DMABN and N-phenylpyrrole molecules (\textcolor{black}{see Tables S15 to S17 and S41 to S43}) should be considered with care. Indeed, we have seen when investigating the PP molecule that tuning the CT character of electronic transitions for a given molecule has an impact on the reliability of the PA-approximation ($d_6$ diagnosis).\\
\indent The PP molecule excited states, when computed at the B3LYP/6-311++G(2d,p) level of theory, exhibit avoided crossing between the first and second singlet excited states and a conical intersection between the second and third excited states. This behaviour is very well reproduced by both the $\phi _S$ and $\phi _S ^\ell$ computation (see the left part of figure \ref{fig:b-c_comparison}), which lead us to state that one should not systematically discard $\phi _S ^\ell$ results for the only reason that it is characterized by a sizeable gradient. However, one sees in figure \ref{fig:b-c_comparison} that when the $\phi _S^\ell$ value drops below 0.4 the values might or might not be reliable. For instance the first electronic transition at this level of theory is quite well described by $\phi _S^\ell$ when tuning down the twist angle from 90 to ca. 55 degrees while the trends completely differ afterward, i.e., when $\phi _S^\ell$ is dropping below 0.4. Significant divergences also occur for the second and third excited state as long as $\phi _S^\ell$ is maintained below 0.4, while non-negligible but tolerable deviations are observed when $\phi _S^\ell$ is lying between 0.4 and 0.5 for the second excited state. \\
\indent Whilst there is neither avoided crossings nor conical intersections in the PP excited-state energy landscape for the first three excited states when computed at the CAM-B3LYP/6-311++G(2d,p) level of theory, the observations reported for the B3LYP xc-functional strictly apply to the CAM-B3LYP computations (see the right part of figure \ref{fig:b-c_comparison}). Finally, the results in Table \ref{tableau 4} show that at given geometries, since the xc-functional influences the CT character, it indirectly influences the accuracy of the PA-approximation. Indeed, while the xc-functional does not explicitly play any role in the definition of the $F$ rule, since we established that the $\phi _S^\ell$ approximation reliability was CT nature-dependent, overtuning the detachment/attachment overlap down by changing the xc-functional at certain geometries might result in $\phi _S$ values that cannot be approximated by Löwdin' symmetrically-orthogonalized orbitals population analysis. Note however that when turning to the $d_8$ diagnosis the conclusions related to the CAM-B3LYP computation of PP excited states are transferrable to the other two range-separated xc-functionals, LC-$\omega$HPBE and $\omega$B97X-D (as well as the CIS and TDHF results). On the other hand the B3LYP results also hold for the other hybrid xc-functional PBE0 while M06-2X reproduces the electronic-structure analysis of range-separated hybrid xc-functionals, which is a known behaviour already reported in Ref. \cite{etienne_fluorene-imidazole_2016}. All the data for the six xc-functionals and the data related to the CIS and TDHF excited-state computations can be found in \textcolor{black}{Tables S51 to S170 and figures S3 to S378}. B3LYP and CAM-B3LYP data are also reported in Table \ref{tableau 4} where two PP twist angles, 80$^\circ$ and 0$^\circ$, have been selected, as they allow us to highlight the particular behaviour of B3LYP for this molecule with a low dihedral angle. \\
\indent Finally, as it is reported in Table \ref{tableau 4} for six basis sets, and in SI for the nine others that were used in this investigation ($d_7$), again, once the convergence on the basis set is achieved for the $\phi _S$ value, the accuracy of the $\phi _S^\ell$ approximation is stable. This allows us to conclude that if the BS used is reliable for the excited-state computation, its size will not influence the accuracy of the $\phi _S^\ell$ approximation. 

\section{Discussion}

\noindent Our approximation here is to consider the Löwdin symmetrically-orthogonalized orbitals as being centered in one point and multiplying or subtracting diagonal entries of $\textbf{S}^{1/2}\textbf{D}\textbf{S}^{1/2}$ and $\textbf{S}^{1/2}\textbf{A}\textbf{S}^{1/2}$ as if we were manipulating point charges with precise locations (centered on the atomic positions), so that Löwdin' symmetric orthogonalization tails are ignored in such a model. However, as we saw, there are few cases in which our approximation is limited. \\
\indent In a diatomic molecule for example: If the atomic orbitals are indeed centered on atomic positions, the orthogonalization tail for the transformation of the orbital centered on the first atom, will be centered on the position of the second one (see figure in chapter 3 of Ref. \cite{mayer_simple_2003}). More generally, Löwdin' symmetrically-orthogonalized orbitals have orthogonalization tails centered on the position of other atoms, and can no longer be considered as strictly centered on one position, unlike the original atomic orbitals. \\
\indent Therefore, for small molecular systems, the presence of these orthogonalization tails introduces a bias in the picture of the electronic distribution given by the population analysis we wish to use. Since the number of atoms is reduced, amongst the atomic orbitals there is a large percentage of them that is susceptible of having a significant overlap, and the orthogonalization tails are distributed on few atoms in a close region of space, so the deviation from our ``local population{\textquotedblright} model (i.e., ignoring the orthogonalization tails), introduces a bias of significant relative importance in the population analysis. \\
\indent The bias can also be significant for very high values of $\phi _S^\ell$ (greater than 0.90), since in that case there are numerous contributions from many pairs of atomic orbitals that are to be considered, which increases the number and importance of the orthogonalization tails that are ignored in the model, or more generally there is a collective effect observed from the sum of the individual small contributions to the deviation from the exact integral value. \\
\indent For low values of $\phi _S^\ell$ (less than 0.5) the problem is quite different: Such low value for the detachment/attachment Löwdin population overlap means that only few diagonal product amplitudes are non-negligible. Since these contributions are summed, introducing a bias on the value of a few number of entries of small amplitude can introduce a strong deviation on the total outcome of the evaluation of $F$.

\section{Conclusion}

\noindent We reported the study of two computational methods for evaluating two types of integrals related to the nature of molecular electronic transitions. The first type of integral is the spatial overlap between two density functions, while the second type separately sums the positive and negative entries of a difference-density function. The first integral evaluation method diagnosed is the projection of the one-body reduced density matrices corresponding to the density functions in the Euclidean space using Cartesian grids, and discrete summations. The second evaluation path is the population analysis of the related density matrices in the basis of symmetrically-orthogonalized orbitals. The diagnosis is related to two rules ($F$ and $H$) mapping the orbital space to a scalar quantity. \\
\indent Our first conclusion concerns the first integral evaluation method: We observed that the precision of the results of our numerical integrations on Cartesian grids is converged for medium-density grids with six points per bohr for the three directions of space, and that the exactness of the trace of some relevant matrices allows the quality of the grid to be diagnosed. \\
\indent When the diagnosis of the population analysis is concerned, we first noticed that the behaviour of the $F$ and $H$ rules was very similar, and decided to focus the analysis on the $F$ rule leading to the Löwdin PA approximation ($\phi _S^\ell$) to the detachment/attachment spatial overlap integral $\phi _S$. \\
\indent We then deduced from an extensive diagnosis of the $F$ rule that its accuracy depends on the size of the target molecular systems, and on the nature of the electronic transitions considered. Basically, the PA-approximation to $\phi _S$ is limited to medium- and large-sized molecular systems. In addition to that, $\phi _S^\ell$ values below 0.5 (for small and medium-sized molecules) and above 0.9 for any type of molecule should not be trusted, so that numerical integrations should be performed in such occurrences. \\
\indent We also noticed that large $\phi _S^\ell$ gradients are not necessarily corresponding to large errors in the population analysis, and that globally the accuracy of the PA-approximation is independent from the size of the basis set used, apart for very small basis sets that in any case would not be used for the accurate computation of excited-state properties in general.

\section*{Supplementary Information}

\noindent Tables S1 to S170, and Figures S1 to S378, are available at the \textcolor{blue}{arxiv.org/abs/1902.05840} address using the ``Ancillary files{\textquotedblright} section.

%

\newpage

\onecolumngrid

\section{Tables and Figures}

\begin{center}
\begin{table}[h!]
\begin{tabular}{lc|cccc}
\hline
Diagnosis & Molecule & System size & CT nature & BS &  xc-functional \\
\hline
$d_1$ & nIII, nV-X, nVI-c & $\surd$ & $\surd$ & $\surd$ &   $\times$ \\
$d_2$ & nV-X & $\times$ & $\surd$ & $\surd$ &   $\times$ \\
\hline
$d_3$ & N$_2$, CO, HCl, H$_2$CO & $\times$ & $\times$ & $\times$ &   $\surd$ \\
\hline
$d_4$ & \multirow{2}{1.5cm}{acenes} & $\surd$ & $\times$ & $\surd$ &   $\surd$\\
$d_5$ & & $\times$ & $\times$ & $\surd$ &   $\times$\\
\hline
$d_6$ & \multirow{3}{1.5cm}{PP} & $\times$ & $\surd$ & $\times$ &   $\times$ \\
$d_7$ & & $\times$ & $\times$ & $\surd$ &   $\times$\\
$d_8$ & & $\times$ & ($\surd$) & $\times$ &   $\surd$\\
\hline
\end{tabular}
\caption{Summary of the eight detachment/attachment PA diagnosis strategies. Molecules for $d_1$ and $d_2$ are taken from Ref. \cite{etienne_toward_2014}; molecules for $d_3$ to $d_5$ are taken from Ref. \cite{peach_excitation_2008}; and PP is taken from Ref. \cite{pastore_unveiling_2017}. The sketches of the molecules are reported in figure S1 of the Supplementary Information.}
\label{tableau 1}
\end{table}
\begin{table}[h!]
\begin{tabular}{cc|cccccc}
\hline
	$n$ & E-S & $\phi_S$ & $\Delta$($\phi_S$) & $\varphi$ & $\Delta$($\varphi$) & $\psi$ & $\Delta$($\psi$) \\
\hline
	\multirow{3}{1.5cm}{1} & 1 & 0.92 & -0.06 & 0.26 & 0.13 & 0.82 & -0.09 \\
	& 2 & 0.98 & 0.00 & 0.15 & 0.00 & 0.90 & 0.00 \\
	& 3 & 0.88 & -0.07 & 0.32 & 0.11 & 0.78 & -0.08 \\
\hline
	\multirow{3}{1.5cm}{2} & 1 & 0.90 & 0.08 & 0.29 & 0.17 & 0.80 & -0.12 \\
	& 2 & 0.98 & -0.01 & 0.15 & 0.02 & 0.90 & -0.01 \\
	& 3 & 0.91 & -0.06 & 0.27 & 0.09 & 0.81 & -0.07 \\
\hline
	\multirow{3}{1.5cm}{3} & 1 & 0.90 & -0.09 & 0.27 & 0.10 & 0.81 & -0.12 \\
	& 2 & 0.97 & -0.01 & 0.15 & 0.03 & 0.90 & -0.02 \\
	& 3 & 0.90 & -0.06 & 0.30 & 0.07 & 0.79 & -0.05 \\
\hline
	\multirow{3}{1.5cm}{4} & 1 & 0.90 & 0.10 & 0.27 & 0.18 & 0.81 & -0.13 \\
	& 2 & 0.89 & -0.05 & 0.30 & 0.05 & 0.79 & -0.04 \\
	& 3 & 0.97 & -0.01 & 0.15 & 0.03 & 0.90 & -0.02 \\
\hline
	\multirow{3}{1.5cm}{5} & 1 & 0.90 & -0.09 & 0.26 & 0.17 & 0.82 & -0.12 \\
	& 2 & 0.90 & -0.06 & 0.33 & 0.06 & 0.78 & -0.05 \\
	& 3 & 0.89 & -0.05 & 0.35 & 0.09 & 0.76 & -0.07 \\
\hline
\end{tabular}
\caption{Descriptors values and their PA-approximated deviations, $\Delta$($q$)$=q$(NI$)-q$(PA), obtained for the three first excited states of the oligo-acenes series with the B3LYP/cc-pVTZ level of theory (part of $d_4$).}
\label{tableau 2}
\end{table}

\begin{table}[h!]
\begin{tabular}{cc|cccccccc}
\hline
	$n$ & E-S & $\phi_S$(NH)& $\Delta$(NH) & $\phi_S$(O) & $\Delta$(O) & $\phi_S$(S) & $\Delta$(S) & $\phi_S$(Se) & $\Delta$(Se) \\
\hline
	\multirow{3}{1.5cm}{1} & 1 & 0.69 & 0.02 & 0.68 & 0.03 & 0.70 & 0.03 & 0.72 & 0.04 \\ 
	& 2 & 0.48 & 0.36 & 0.48 & 0.36 & 0.48 & 0.39 & 0.48 & 0.38 \\
	& 3 & 0.34 & 0.14 & 0.52 & 0.40 & 0.51 & 0.41 & 0.54 & 0.03 \\
\hline
	\multirow{3}{1.5cm}{2} & 1 & 0.57 & 0.01 & 0.60 & 0.03 & 0.58 & 0.02 & 0.68 & 0.02 \\
	& 2 & 0.49 & 0.42 & 0.48 & 0.35 & 0.76 & 0.25 & 0.52 & 0.25 \\
	& 3 & 0.40 & 0.03 & 0.71 & 0.04 & 0.72 & 0.00 & 0.60 & 0.12 \\
\hline
	\multirow{3}{1.5cm}{3} & 1 & 0.48 & 0.01 & 0.53 & 0.02 & 0.51 & 0.02 & 0.62 & 0.02 \\
	& 2 & 0.73 & -0.01 & 0.78 & -0.02 & 0.75 & -0.01 & 0.79 & -0.01 \\
	& 3 & 0.63 & 0.01 & 0.69 & 0.00 & 0.70 & -0.01 & 0.74 & -0.01 \\
\hline
	\multirow{3}{1.5cm}{4} & 1 & 0.39 & 0.00 & 0.47 & 0.02 & 0.52 & 0.01 & 0.57 & 0.01 \\
	& 2 & 0.60 & 0.00 & 0.77 & -0.02 & 0.75 & -0.02 & 0.78 & -0.01 \\
	& 3 & 0.71 & -0.03 & 0.66 & 0.00 & 0.70 & -0.01 & 0.74 & -0.01 \\
\hline
	\multirow{3}{1.5cm}{5} & 1 & 0.34 & 0.01 & 0.42 & 0.01 & 0.47 & 0.00 & 0.53 & 0.00 \\
	& 2 & 0.52 & 0.00 & 0.75 & -0.02 & 0.73 & -0.02 & 0.76 & -0.02 \\
	& 3 & 0.71 & -0.04 & 0.65 & -0.01 & 0.68 & -0.02 & 0.72 & -0.02 \\
\hline
\end{tabular}
\caption{Detachment/attachment spatial overlap integral values and their PA-approximated deviations, $\Delta$(X)$=\phi_S$(X)$-\phi_S^{\ell}$(X), calculated for the different molecules from the nV-X molecular test-set (part of $d_1$). Those values were obtained using PBE0/6-311++G(2d,p) excited-state calculations, based on PBE0/6-311G(d,p) geometries.}
\label{tableau 3}
\end{table}
\begin{table}[h!]
\begin{tabular}{ll|cccc|cccc|cccc}
\hline
\multicolumn{13}{c}{angle 0$^\circ$} \\
\hline
	\multirow{2}{0.0cm}{BS} & \multirow{2}{0.5cm}{\textit{K}} & \multicolumn{2}{c}{B3LYP} & \multicolumn{2}{c|}{CAM-B3LYP} & \multicolumn{2}{c}{B3LYP} & \multicolumn{2}{c|}{CAM-B3LYP} & \multicolumn{2}{c}{B3LYP} & \multicolumn{2}{c}{CAM-B3LYP}  \\
	& & $\phi_S$ & $\phi_S^{\ell}$ & $\phi_S$ & $\phi_S^{\ell}$ & $\varphi$ & $\varphi^{\ell}$ & $\varphi$ & $\varphi^{\ell}$ & $\psi$ & $\psi^{\ell}$ & $\psi$ & $\psi^{\ell}$ \\
\hline
	3-21G & 135 & 0.33 & 0.43 & 0.74 & 0.71 & 0.86 & 0.82 & 0.60 & 0.63 & 0.23 & 0.31 & 0.57 & 0.54 \\
	6-31G & 135 & 0.32 & 0.42 & 0.73 & 0.69 & 0.86 & 0.83 & 0.62 & 0.65 & 0.23 & 0.30 & 0.55 & 0.52 \\
	6-31+G & 187 & 0.31 & 0.41 & 0.70 & 0.66 & 0.87 & 0.83 & 0.63 & 0.66 & 0.22 & 0.30 & 0.53 & 0.50 \\
	6-31G(d) & 213 & 0.31 & 0.42 & 0.73 & 0.69 & 0.86 & 0.83 & 0.61 & 0.65 & 0.22 & 0.30 & 0.56 & 0.52 \\
	6-311+G(d) & 313 & 0.31 & 0.40 & 0.71 & 0.66 & 0.87 & 0.84 & 0.63 & 0.66 & 0.22 & 0.29 & 0.54 & 0.50 \\
	6-311++G(2d,p) & 414 & 0.31 & 0.41 & 0.70 & 0.66 & 0.87 & 0.83 & 0.63 & 0.66 & 0.22 & 0.29 & 0.53 & 0.50 \\
\hline
\multicolumn{13}{c}{angle 80$^\circ$} \\
\hline
	\multirow{2}{1.5cm}{BS} & \multirow{2}{1.5cm}{\textit{K}} & \multicolumn{2}{c}{B3LYP} & \multicolumn{2}{c|}{CAM-B3LYP} & \multicolumn{2}{c}{B3LYP} & \multicolumn{2}{c|}{CAM-B3LYP} & \multicolumn{2}{c}{B3LYP} & \multicolumn{2}{c}{CAM-B3LYP}  \\
	& & $\phi_S$ & $\phi_S^{\ell}$ & $\phi_S$ & $\phi_S^{\ell}$ & $\varphi$ & $\varphi^{\ell}$ & $\varphi$ & $\varphi^{\ell}$ & $\psi$ & $\psi^{\ell}$ & $\psi$ & $\psi^{\ell}$ \\
\hline
3-21G & 135 & 0.53 & 0.54 & 0.37 & 0.34 & 0.81 & 0.82 & 0.90 & 0.93 & 0.37 & 0.37 & 0.25 & 0.22 \\ 
6-31G & 135 & 0.55 & 0.55 & 0.33 & 0.30 & 0.81 & 0.82 & 0.92 & 0.94 & 0.38 & 0.38 & 0.22 & 0.20 \\ 
6-31+G & 187 & 0.50 & 0.51 & 0.30 & 0.26 & 0.84 & 0.85 & 0.93 & 0.95 & 0.35 & 0.34 & 0.20 & 0.17 \\
6-31G(d) & 213 & 0.54 & 0.54 & 0.33 & 0.30 & 0.81 & 0.83 & 0.92 & 0.94 & 0.37 & 0.37 & 0.22 & 0.20  \\
6-311+G(d) & 313 & 0.49 & 0.50 & 0.30 & 0.26 & 0.84 & 0.86 & 0.93 & 0.95 & 0.34 & 0.33 & 0.20 & 0.17 \\
6-311++G(2d,p) & 414 & 0.49 & 0.50 & 0.30 & 0.27 & 0.84 & 0.85 & 0.93 & 0.94 & 0.34 & 0.33 & 0.20 & 0.18 \\
\hline \\ \end{tabular}
\caption{Descriptor values for the first three excited states of PP obtained with B3LYP and CAM-B3LYP xc-functionnals and six BSs out of the fifteen investigated in this contribution. Two dihedral angles have been choosen (0$^\circ$ and 80$^\circ$). $K$ is the number of basis functions.}
\label{tableau 4}
\end{table}
\end{center}

\vspace*{\fill} \nopagebreak
%
%
%
	
%

\newpage

\setcounter{figure}{0}

\begin{center}
\begin{figure}[h!]
\includegraphics[scale=1]{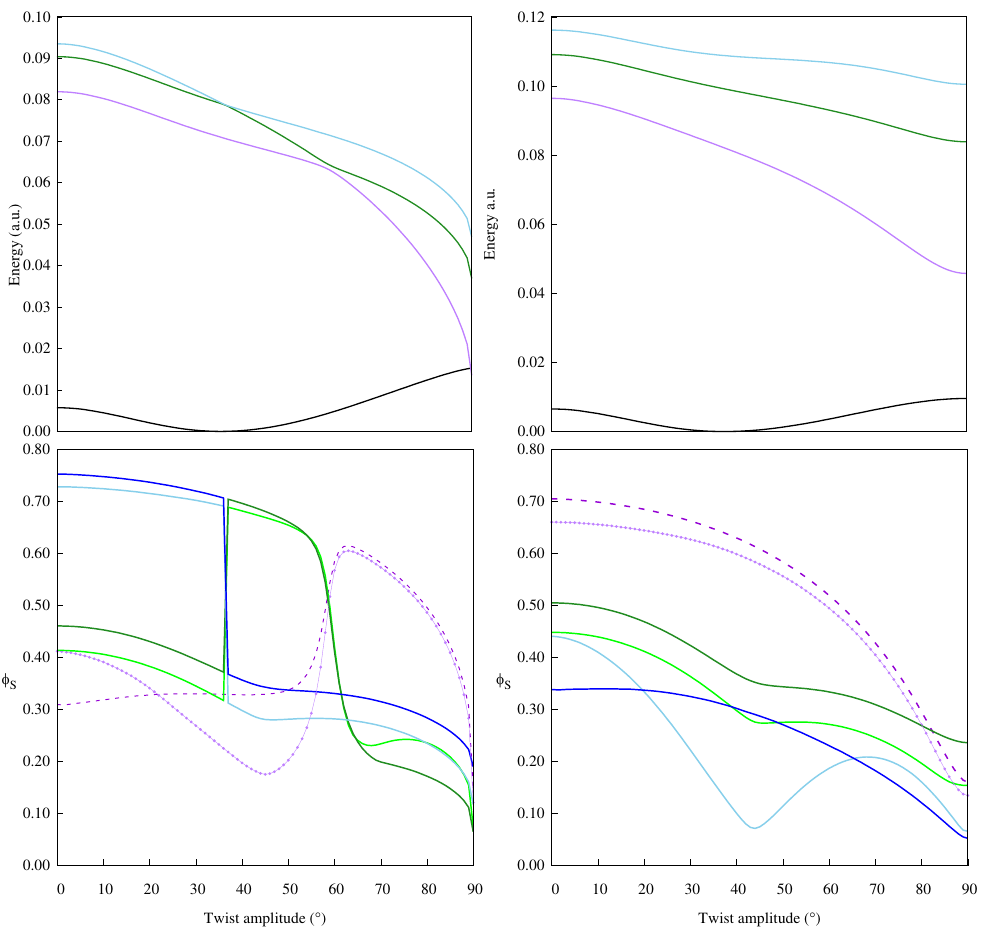}
\caption{(Top) Energy landscapes for the PP molecule along the scan of its central dihedral (twist) angle. The black line is the ground-state energy, the purple one is the first singlet excited-state energy, the green one is the second singlet excited-state energy, and the blue one is the third singlet excited-state energy. The left part is obtained using the B3LYP/6-311++G(2d,p) level of theory, and the right part is obtained using the CAM-B3LYP level of theory. (Bottom) Comparison of the $\phi _S$ and $\phi _S^\ell$ values, with the same color code and levels of theory as the top part of the figure. Deep blue and green colors are for the NI values of the second and third excited states, the dashed line represents the NI values for the first singlet excited state, and the dotted line represents the PA values for the first singlet excited state.}
\label{fig:b-c_comparison}
\end{figure}
\end{center}

\clearpage

\twocolumngrid

\section{Appendices}\label{sec:App}

\subsection{Nomenclature and comments}\label{sub:appNom}

\noindent The modifications related in this Appendix concern the nomenclature of objects, functions, and quantities, taken from Refs. \cite{etienne_toward_2014,etienne_new_2014,etienne_probing_2015,pastore_unveiling_2017,etienne_theoretical_2017,
etienne_charge_2018}. 

Since we are not dealing anymore with centroids of charge, the use of ``$\bm{\xi} = (\xi_1,\xi _2, \xi _3)${\textquotedblright} for indicating the position vector $\textbf{r} = (x,y,z)$ is no longer necessary for writing compact formulas, so any three-dimensional position vector has been denoted as ``$\textbf{r}${\textquotedblright} in this paper. Density functions were denoted with an ``$n${\textquotedblright} here rather than the ``$\varrho${\textquotedblright} used in Refs. \cite{etienne_toward_2014,etienne_new_2014,etienne_probing_2015}.

When detachment and attachment 1--RDMs are used in the atomic space, those are no longer written $\bm{\Gamma}$ and $\bm{\Lambda}$, but simply \textbf{D} and \textbf{A} respectively. The previous nomenclature was used in order to avoid any confusion with the terms ``Donor{\textquotedblright} and ``Acceptor{\textquotedblright}, commonly used in molecular science for talking about fragments that can donate or accept electron(s).

Note also that in the previously cited contributions, some quantities were given a definition corresponding to their practical evaluation, like the $\tilde{\chi}$ and $\tilde{\varphi}$ descriptors, which used to take a $\sim$ when derived from the detachment/attachment density difference, in contrast to their evaluation from the excited/ground state density difference. Though in practice the values might differ due to some numerical imprecision in their computation, which explains the use of this symbol to highlight how the quantity was actually computed, this symbol has been dropped according to the fact that the quantities derived from the two derivation schemes are, by construction, identical in principle.  

Finally, when relaxed 1--DDMs were used in Ref. \cite{pastore_unveiling_2017}, the related quantities were labeled with an ``R{\textquotedblright} while the unrelaxed ones were labeled with a ``U{\textquotedblright}. In this contribution, the ``relaxed{\textquotedblright} quantities were simply labelled with a prime while the unrelaxed ones were bearing no distinctive mark.

\subsection{Correspondence between some descriptors related to \textit{G} and \textit{H}}\label{sub:appGH}

\textit{Around G}

$\;$

\noindent The four density-based descriptors previously derived using a $G$--like functional were $\zeta ^\mathrm{Z}$ (unchanged here), as well as $\tilde{\chi}$, $\tilde{\varphi}$, and $\alpha ^\ddag _{\bm{\nu}}$ with $\bm{\nu} \in \left\lbrace\bm{\Gamma},\bm{\Lambda}\right\rbrace$, renamed respectively ${\chi}$, ${\varphi}$, and $\alpha ^\ddag _{\bm{\nu}}$ with $\bm{\nu} \in \left\lbrace \textbf{D} , \textbf{A}\right\rbrace$ in this contribution. 

$\;$

$\;$

$\;$

$\;$

$\;$

$\;$

\textit{Around H}

$\;$

\noindent The $\tilde{\chi}_1$, $\tilde{\varphi}_1$, and $\phi _S^{\Omega}(1)$  in Ref. \cite{etienne_probing_2015} have been renamed respectively $\chi ^{\ell}$, $\varphi ^{\ell}$, and $\phi _S^\ell$ here, while $\tilde{\chi}_2$, $\tilde{\varphi}_2$ and $\phi _S^\Omega (2)$ are not explicitely mentioned in this paper, but correspond to $\chi ^{(x=0})$, $\varphi ^{(x=0})$, and $\phi_S ^{(x=0)}$ according to the current nomenclature.

\subsection{The particular case of Löwdin population analysis} \label{sub:appLow}

\noindent In this appendix we demonstrate that when generalizing the orbital population analysis, only the Löwdin model prevents us from deriving unphysical atomic orbital populations (negative or greater than the maximal allowed occupancy).

For this purpose we will show that for a two-electron system with a two-orbital model, any other population analysis can, in principle, lead to an unphysical atomic population. In this appendix, any matrix element $(\textbf{L})_{ij}$ will be written $L_{ij}$ for the sake of readability. This applies to the matrix elements of \textbf{C}, \textbf{P}, and \textbf{S}.

Let us consider a basis of two real-valued normalised but not orthogonal atomic orbitals, $\left( \phi _1 , \phi _2 \right)$. Their overlap matrix reads
\begin{eqnarray}
 \textbf{S} =  \left( 
 \begin{array}{cc}
1  & S   \\
S  &  1 \\
 \end{array}\right) \qquad -1 < S < 1.
\end{eqnarray}
If $S$ was allowed to take the $-1$ or the $1$ value, the atomic orbitals would no longer be linearly independent.

Let us now consider two real-valued molecular orbitals, $ \varphi _1$ and $\varphi _2$, that form a complete and orthonormal basis with respect to the space spanned by the atomic orbitals basis set. Their spatial parts, $\chi _1 $ and $\chi _2 $, expand as
\begin{align}
\nonumber \chi _1 (\textbf{r}) &= C_{11} \phi _1 (\textbf{r}) + C_{21} \phi _2 (\textbf{r}) \\
\chi _2 (\textbf{r}) &= C_{12} \phi _1 (\textbf{r}) + C_{22} \phi _2 (\textbf{r}).
\end{align}
The expansion coefficients are real (note that the present derivation can be generalised easily to complex-valued atomic and molecular orbitals). The matrix of coefficients,
\begin{eqnarray}
 \textbf{C} =  \left( 
 \begin{array}{cc}
C_{11}  & C_{12}   \\
C_{21}  &  C_{22} \\
 \end{array}\right) ,
\end{eqnarray}
satisfies a generalised orthonormality relationship with respect to the descriptor defined by \textbf{S}, expressed as
\begin{equation}
\textbf{C}^\dag \textbf{SC} = \textbf{1}.
\end{equation}
The molecular orbitals are assumed to be linearly independent, hence \textbf{C} is invertible and
\begin{equation} \textbf{CC}^\dag = \textbf{S}^{-1}.
\end{equation}
\begin{widetext}
\noindent \textbf{S} is a Gram matrix. Therefore, it is positive definite. Its eigenvalues are $1 \pm S > 0$. A possible eigendecomposition of it is 
\begin{eqnarray}
\textbf{S} = \underbrace{\left( 
\begin{array}{cc}
\dfrac{1}{\sqrt{2}} & \dfrac{-1}{\sqrt{2}} \\
& \\
\dfrac{1}{\sqrt{2}} & \dfrac{1}{\sqrt{2}}
\end{array} \right)}_{\displaystyle \textbf{U}}\overbrace{ \left( 
\begin{array}{cc}
1+S & 0 \\ 0 & 1-S \\
\end{array} \right)}^{\displaystyle\bm{\sigma}} \underbrace{\left(
\begin{array}{cc}
\dfrac{1}{\sqrt{2}} & \dfrac{1}{\sqrt{2}} \\
&   \\
\dfrac{-1}{\sqrt{2}} & \dfrac{1}{\sqrt{2}}
\end{array} \right)}_{\displaystyle \textbf{U}^\dag}.
\end{eqnarray}
From this, we can define, for any real $x$,
\begin{eqnarray}\hspace*{-1cm} \nonumber
\textbf{S}^x = \underbrace{\left( 
\begin{array}{cc}
\dfrac{1}{\sqrt{2}} & \dfrac{-1}{\sqrt{2}} \\
&  \\
\dfrac{1}{\sqrt{2}} & \dfrac{1}{\sqrt{2}}
\end{array} \right)}_{\displaystyle \textbf{U}}\overbrace{ \left( 
\begin{array}{cc}
(1+S)^x & 0 \\ 0 & (1-S)^x \\
\end{array} \right)}^{\displaystyle\bm{\sigma}^x} \underbrace{\left(
\begin{array}{cc}
\dfrac{1}{\sqrt{2}} & \dfrac{1}{\sqrt{2}} \\
&   \\
\dfrac{-1}{\sqrt{2}} & \dfrac{1}{\sqrt{2}}
\end{array} \right)}_{\displaystyle \textbf{U}^\dag} = \left(
\begin{array}{cc}
\dfrac{(1+S)^x + (1-S)^x}{{2}} & \dfrac{(1+S)^x - (1-S)^x}{{2}} \\
&   \\
\dfrac{(1+S)^x - (1-S)^x}{{2}} & \dfrac{(1+S)^x + (1-S)^x}{{2}}
\end{array} \right)
\end{eqnarray}
where $(1\pm S)^x > 0$. In particular,
\begin{eqnarray}
\textbf{S}^{1/2} = \left(
\begin{array}{cc}
\dfrac{\sqrt{(1+S)} + \sqrt{(1-S)}}{{2}} & \dfrac{\sqrt{(1+S)} - \sqrt{(1-S)}}{{2}} \\
& \\
\dfrac{\sqrt{(1+S)} - \sqrt{(1-S)}}{{2}} & \dfrac{\sqrt{(1+S)} + \sqrt{(1-S)}}{{2}}
\end{array} \right),
\end{eqnarray}
\begin{eqnarray}
\textbf{S}^{-1} = \dfrac{1}{1-S^2}\left(
\begin{array}{cc}
1 & -S \\
-S & 1
\end{array} \right),
\end{eqnarray}
and
\begin{eqnarray}
\textbf{S}^{-1/2} = \dfrac{1}{\sqrt{1-S^2}}\left(
\begin{array}{cc}
\dfrac{\sqrt{(1+S)} + \sqrt{(1-S)}}{{2}} & \dfrac{\sqrt{(1-S)} - \sqrt{(1+S)}}{{2}} \\
&  \\
\dfrac{\sqrt{(1-S)} - \sqrt{(1+S)}}{{2}} & \dfrac{\sqrt{(1+S)} + \sqrt{(1-S)}}{{2}}
\end{array} \right).
\end{eqnarray}
Among the possible square roots of \textbf{S}, the only positive definite one, which is unique and termed {\textquotedblleft}principal{\textquotedblright} square root of \textbf{S}, is chosen here, and written $\textbf{S}^{1/2}$.

The generalised orthonormality relationship fulfilled by \textbf{C} can be recast as follows (Löwdin symmetric orthonormalisation),
\begin{equation}
\left(\textbf{S}^{1/2}\textbf{C}\right)^\dag \left(\textbf{S}^{1/2}\textbf{C}\right) = \textbf{1}.
\end{equation}
Thus, \textbf{C} being square and invertible,
\begin{equation}
 \left(\textbf{S}^{1/2}\textbf{C}\right)\left(\textbf{S}^{1/2}\textbf{C}\right)^\dag = \textbf{1} \Rightarrow \left(\textbf{S}^{1/2}\textbf{C}\right)^\dag = \left(\textbf{S}^{1/2}\textbf{C}\right)^{-1}.
\end{equation}
Note that this would not hold if the molecular orbitals set were not complete (i.e., if \textbf{C} were rectangular). $\textbf{S}^{1/2}\textbf{C}$ is a real unitary matrix (i.e., orthogonal) and can be chosen as a rotation matrix (this parametrisation is not unique),
\begin{eqnarray}
\textbf{S}^{1/2}\textbf{C} = \textbf{R} = \left(
\begin{array}{cc}
\mathrm{cos} \theta & - \mathrm{sin} \theta \\
\mathrm{sin} \theta & \mathrm{cos} \theta
\end{array} \right),
\end{eqnarray}
where $\theta$ is real. As a consequence,
\begin{equation}
\textbf{C} = \textbf{S}^{-1/2}\textbf{R},
\end{equation}
so that 
\begin{eqnarray} \hspace*{-1cm}\nonumber
\textbf{C} = \dfrac{1}{\sqrt{1-S^2}} \left(
\begin{array}{cc}
\dfrac{\sqrt{1+S}(\mathrm{cos} \theta - \mathrm{sin} \theta) + \sqrt{1-S}(\mathrm{cos} \theta + \mathrm{sin} \theta)}{2} & \dfrac{\sqrt{1-S}(\mathrm{cos} \theta - \mathrm{sin} \theta) - \sqrt{1+S}(\mathrm{cos} \theta + \mathrm{sin} \theta)}{2} \\ & \\
\dfrac{-\sqrt{1+S}(\mathrm{cos} \theta - \mathrm{sin} \theta) + \sqrt{1-S}(\mathrm{cos} \theta + \mathrm{sin} \theta)}{2} & \dfrac{\sqrt{1-S}(\mathrm{cos} \theta - \mathrm{sin} \theta) + \sqrt{1+S}(\mathrm{cos} \theta + \mathrm{sin} \theta)}{2} 
\end{array}\right),
\end{eqnarray}
\end{widetext}
Let us now consider a two-electron closed-shell wave function in the form a single Slater determinant,
\begin{equation}
\Psi (\textbf{s}_1,\textbf{s}_2) = \left| \varphi _1 (\textbf{s}_1) \bar{\varphi}_1 (\textbf{s}_2)\right|
\end{equation}
where ``\textbf{s}{\textquotedblright} collects the four spin-spatial variables. The corresponding 1--CD reads
\begin{equation}
n(\textbf{r}) = 2 \left| \chi _1 (\textbf{r}) \right| ^2
\end{equation}
where, again, $\chi_1$ is the spatial part of $\varphi _1$. The $n$ function satisfies
\begin{align}\nonumber
n(\textbf{r}) &= 2(C_{11}^2 |\phi _1(\textbf{r})|^2 + 2C_{11}C_{21} \phi _1 (\textbf{r}) \phi _2 (\textbf{r}) + C_{21}^2 | \phi _2 (\textbf{r})|^2) \\ &= \sum _{\sigma=1,2}\sum _{\lambda=1,2} P_{\sigma \lambda} \phi _\sigma (\textbf{r}) \phi _\lambda (\textbf{r}).
\end{align}
The corresponding density matrices, \textbf{P} and the duodempotent $\bm{\gamma}$, satisfy
\begin{equation}
P_{\sigma\lambda} = 2 C_{\sigma1} C_{\lambda1} ,
\end{equation}
\begin{equation}\bm{\gamma} = 2 \oplus 0.
\end{equation}
The Mulliken gross orbital populations are defined as 
\begin{equation}
g_\sigma = \sum _{\lambda = 1,2} P_{\sigma \lambda} S_{\lambda \sigma} = (\textbf{PS})_{\sigma\sigma}.
\end{equation}
Thus, 
\begin{equation}
\sum _{\sigma \in \left\lbrace 1,2\right\rbrace} g_\sigma = \int _{\mathbb{R}^3} d\textbf{r}\, n(\textbf{r}) = 2.
\end{equation}
\begin{widetext}
\noindent From the expression of $\bm{\gamma}$ and \textbf{C} given above, and according to Eq. \eqref{eq:Pgamma} we get
\begin{eqnarray}\nonumber
\textbf{P} = 2\left(
\begin{array}{cc}
C_{11}^2 & C_{11}C_{21} \\ & \\
C_{11}C_{21} & C_{21}^2
\end{array}\right) = \dfrac{1}{1-S^2} \left(
\begin{array}{cc}
1-S\mathrm{sin} (2\theta)+\sqrt{1-S^2}\mathrm{cos}(2\theta) & \mathrm{sin}(2\theta) - S \\ & \\
\mathrm{sin}(2\theta) -S & 1-S \mathrm{sin}(2\theta) - \sqrt{1-S^2}\mathrm{cos}(2\theta) 
\end{array}\right),
\end{eqnarray}
\begin{eqnarray}\nonumber
\textbf{PS} = 2 \left(
\begin{array}{cc}
C_{11}^2 + S C_{11}C_{21} & C_{11}C_{21} + SC_{11}^2 \\
&   \\
C_{11}C_{21} + SC_{21}^2 & C_{21}^2 + SC_{11}C_{21}
\end{array}\right) = \left(
\begin{array}{cc}
1+ \dfrac{\mathrm{cos}(2\theta)}{\sqrt{1-S^2}} & \mathrm{sin}(2\theta) + S \dfrac{\mathrm{cos}(2\theta)}{\sqrt{1-S^2}} \\
&  \\
\mathrm{sin}(2\theta) - S \dfrac{\mathrm{cos}(2\theta)}{\sqrt{1-S^2}} & 1-\dfrac{\mathrm{cos}(2\theta)}{\sqrt{1-S^2}}
\end{array}\right). \nonumber
\end{eqnarray}
This can be formulated using expansions in terms of Pauli matrices
\begin{eqnarray}
\bm{\sigma}_1 = \left(
\begin{array}{cc}
0 & 1 \\ 1 & 0
\end{array}\right) \qquad 
\bm{\sigma}_2 = \left(
\begin{array}{cc}
0 & -i \\ i & 0
\end{array}\right) \qquad
\bm{\sigma}_3 = \left(
\begin{array}{cc}
1 & 0 \\ 0 & -1
\end{array}\right)  
\end{eqnarray}
as
\begin{align} \nonumber 
\textbf{P} &= 2 \left( \dfrac{C_{11}^2 + C_{21}^2}{2}\textbf{1} + C_{11}C_{21}\bm{\sigma}_1 + \dfrac{C_{11} ^2 - C_{21}^2}{2}\bm{\sigma}_3 \right)   \\ &= \dfrac{1}{1-S^2} \left(  [1-S\mathrm{sin}(2\theta)]\textbf{1}  +  [\mathrm{sin}(2\theta) - S] \bm{\sigma}_1  +  \left[\sqrt{1-S^2}\mathrm{cos}(2\theta) \right] \bm{\sigma}_3\right)\nonumber 
\end{align}
where we identify
\begin{equation}
P_0 = \dfrac{1-S\mathrm{sin}(2\theta)}{1-S^2}  ; \qquad P_1 = \dfrac{\mathrm{sin}(2\theta) - S}{1-S^2}  ; \qquad P_3 = \dfrac{\sqrt{1-S^2}\mathrm{cos}(2\theta)}{1-S^2}.
\end{equation}
We can also rewrite
\begin{equation}
\textbf{S} = \textbf{1} + S\bm{\sigma}_1 ; \qquad \textbf{PS} = \textbf{1} + \mathrm{sin}(2\theta) \bm{\sigma}_1 + S \dfrac{\mathrm{cos}(2\theta)}{\sqrt{1-S^2}}i \bm{\sigma}_2 + \dfrac{\mathrm{cos}(2\theta)}{\sqrt{1-S^2}} \bm{\sigma}_3.
\end{equation}
Generalised orbital-population analyses can also be considered, based on $\textbf{S}^x\textbf{PS}^{1-x}$ with $0 \leq x \leq 1$. We get
\begin{equation}
\textbf{S}^x = \underbrace{\dfrac{(1+S)^x + (1-S)^x}{2}}_{\displaystyle S_0^{(x)}}\textbf{1} + \underbrace{\dfrac{(1+S)^x - (1-S)^x}{2}}_{\displaystyle S_1^{(x)}}\bm{\sigma}_1,
\end{equation}
\begin{align*}
\textbf{S}^x\textbf{PS}^{1-x} &= \nonumber \left( S_0^{x}\textbf{1} + S_1^{(x)}\bm{\sigma}_1\right) \left( P_0 \textbf{1} + P_1 \bm{\sigma}_1 + P_3 \bm{\sigma}_3\right)\left(S_0^{(1-x)}\textbf{1}+S_1^{(1-x)}\bm{\sigma}_1\right) \\
&= \nonumber \left[ \left(S_0^{(x)}S_0^{(1-x)}+S_1^{(x)}S_1^{(1-x)}\right)P_0 + \left( S_0^{(x)}S_1^{(1-x)}+S_1^{(x)}S_0^{(1-x)} \right) P_1\right] \textbf{1} \\
&+ \nonumber \left[ \left(S_0^{(x)}S_1^{(1-x)}+S_1^{(x)}S_0^{(1-x)}\right)P_0 + \left( S_0^{(x)}S_0^{(1-x)}+S_1^{(x)}S_1^{(1-x)}\right) P_1\right] \bm{\sigma}_1  \\
&+   \left(S_0^{(x)}S_1^{(1-x)}-S_1^{(x)}S_0^{(1-x)}\right)P_3 i \bm{\sigma}_2 + \left( S_0^{(x)}S_0^{(1-x)}-S_1^{(x)}S_1^{(1-x)}\right) P_3 \bm{\sigma}_3 ,
\end{align*}
with
\begin{align*}
S_0^{(x)}S_0^{(1-x)} &= \dfrac{2+(1-S)^x(1+S)^{1-x}+(1+S)^x(1-S)^{1-x}}{4}, \\
S_0^{(x)}S_1^{(1-x)} &= \dfrac{2S+(1-S)^x(1+S)^{1-x}-(1+S)^x(1-S)^{1-x}}{4}, \\
S_1^{(x)}S_0^{(1-x)} &= \dfrac{2S-(1-S)^x(1+S)^{1-x}+(1+S)^x(1-S)^{1-x}}{4}, \\
S_1^{(x)}S_1^{(1-x)} &= \dfrac{2-(1-S)^x(1+S)^{1-x}-(1+S)^x(1-S)^{1-x}}{4}, \\
S_0^{(x)}S_0^{(1-x)} + S_1^{(x)}S_1^{(1-x)} &= 1, \\
S_0^{(x)}S_1^{(1-x)} + S_1^{(x)}S_0^{(1-x)} &= S, \\
S_0^{(x)}S_0^{(1-x)} - S_1^{(x)}S_1^{(1-x)} &= \dfrac{(1-S)^x(1+S)^{1-x}+(1+S)^x(1-S)^{1-x}}{2}, \\
P_0+SP_1 &= C_{11}^2 + C_{21}^2 + 2SC_{11}C_{21} = \dfrac{1-S \mathrm{sin}(2\theta)}{1-S^2} + S\dfrac{\mathrm{sin}(2\theta) - S}{1-S^2} = 1.
\end{align*}
The diagonal entries (orbital populations), thus read
\begin{align*}
(\textbf{S}^x\textbf{PS}^{1-x})_{11} = (\textbf{S}^{1-x}\textbf{PS}^{x})_{11} &=
P_0+SP_1 + \dfrac{(1-S)^x(1+S)^{1-x}+(1+S)^x(1-S)^{1-x}}{2}P_3 \\
&= 1+ \dfrac{(1-S)^x(1+S)^{1-x}+(1+S)^x(1-S)^{1-x}}{2}\left(C_{11}^2 - C_{21}^2\right) \\
&= 1+ \dfrac{(1-S)^x(1+S)^{1-x}+(1+S)^x(1-S)^{1-x}}{2}\dfrac{\mathrm{cos}(2\theta)}{\sqrt{1-S^2}}
\end{align*}
and
\begin{align*}
(\textbf{S}^x\textbf{PS}^{1-x})_{22} = (\textbf{S}^{1-x}\textbf{PS}^{x})_{22} &=
P_0+SP_1 - \dfrac{(1-S)^x(1+S)^{1-x}+(1+S)^x(1-S)^{1-x}}{2}P_3 \\
&= 1- \dfrac{(1-S)^x(1+S)^{1-x}+(1+S)^x(1-S)^{1-x}}{2}\left(C_{11}^2 - C_{21}^2\right) \\
&= 1- \dfrac{(1-S)^x(1+S)^{1-x}+(1+S)^x(1-S)^{1-x}}{2}\dfrac{\mathrm{cos}(2\theta)}{\sqrt{1-S^2}}.
\end{align*}
For $x=0$ or $x=1$ (Mulliken), the populations satisfy
\begin{align*}
1-\dfrac{1}{\sqrt{1-S^2}} \leq (\textbf{PS})_{11} = (\textbf{SP})_{11} &= 1+\dfrac{\mathrm{cos}(2\theta)}{\sqrt{1-S^2}} \leq 1+ \dfrac{1}{\sqrt{1-S^2}} \\
1-\dfrac{1}{\sqrt{1-S^2}} \leq (\textbf{PS})_{22} = (\textbf{SP})_{22} &= 1-\dfrac{\mathrm{cos}(2\theta)}{\sqrt{1-S^2}} \leq 1+ \dfrac{1}{\sqrt{1-S^2}}.
\end{align*}
They can be negative, depending on the values of $S$ and $\theta$. Large values of $|S|$ make this situation likely to occur (see the top of figure S2). It might also happen that the orbital population exceeds the maximal allowed orbital population (see the bottom of figure S2). These two situations can occur also for any other value of $0 \leq x \leq 1$ except for $x=1/2$. In the latter case, $\textbf{S}^{1/2}\textbf{PS}^{1/2}$ is positive semidefinite and its diagonal entries are all nonnegative. In particular here,
\begin{align*}
0 \leq \left(\textbf{S}^{1/2}\textbf{PS}^{1/2}\right)_{11} = 1+ \mathrm{cos}(2\theta) \leq 2,\\
0 \leq \left(\textbf{S}^{1/2}\textbf{PS}^{1/2}\right)_{22} = 1- \mathrm{cos}(2\theta) \leq 2.
\end{align*}
This justifies that only Löwdin-like population analysis was used in our calculations.
\end{widetext}

\bibliographystyle{ieeetr}

\end{document}